\documentclass[twocolumn,prb]{revtex4}

\usepackage{epsfig}
\usepackage{dcolumn}
\usepackage{bm}

\begin{document}

\title{Josephson Effect in Pb/I/NbSe$_2$ Scanning Tunneling Microscope Junctions}

\author{O.~Naaman}
\author{R.~C.~Dynes}
\affiliation{Department of Physics, University of California, San
Diego\\ 9500 Gilman Drive
La Jolla, California 92093}

\author{E.~Bucher}

\affiliation{Lucent Technologies, Murray Hill, New Jersey 07974 }


\begin{abstract}
We have developed a method for the reproducible fabrication of
superconducting scanning tunneling microscope (STM) tips. We use
these tips to form superconductor/insulator/superconductor tunnel
junctions with the STM tip as one of the electrodes. We show that
such junctions exhibit fluctuation dominated Josephson effects,
and describe how the Josephson product I$_c$R$_N$ can be inferred
from the junctions' tunneling characteristics in this regime.
This is first demonstrated for tunneling into Pb films, and then
applied in studies of single crystals of NbSe$_2$. We find that
in NbSe$_2$, I$_c$R$_N$ is lower than expected, which could be
attributed to the interplay between superconductivity and the
coexisting charge density wave in this material.
\end{abstract}

\pacs{Valid PACS appear here}
\maketitle

\section{Introduction}

Scanning tunneling microscopy (STM) has been proven to be an
invaluable tool in the study of superconductors by serving as a
probe of the density of states with very high energy resolution
and sub-nanometer spatial
resolution\cite{Hess90,Yazdani97,Pan00,Hoffman02}. STM has also
been successfully used to create tunnel junctions on materials
that do not readily form high quality planar junctions ({\it
e.g.~}in MgB$_2$)\cite{Rubio01}, or in cases where a particular
small region of the sample is of interest\cite{Truscott99}.

In recent years it has been
realized\cite{Pan98,Naaman01a,Smakov01} that the use of
superconducting (SC) STM tips may provide a complementary
technique to conventional STM studies, since in SC STM both
quasiparticle and Josephson tunneling can be measured locally
with a scannable superconductor/insulator/superconductor (S/I/S)
junction. Here we report on experiments we performed using SC STM
tips. We measure the tunneling characteristics of Pb and NbSe$_2$
samples at T$\sim$2 K for various values of the junctions' normal
state resistances. For junction resistances below 100 k$\Omega$
we observe fluctuation dominated Josephson effects, and estimate
the Josephson I$_c$R$_N$ product, a physically important quantity,
for tunneling into NbSe$_2$.

\section{Experimental Technique}

\begin{figure}
\epsfxsize=\columnwidth \epsfysize=3.0in
\epsfbox{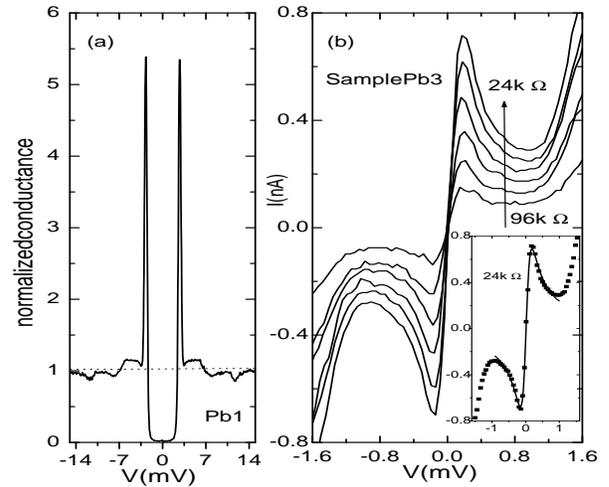} \caption{\label{fig1}(a) dI/dV curve
for high resistance junction between a Pb/Ag tip and a Pb/Ag
sample at 2.1 K. The Pb phonon structure is clearly observed. (b)
$I-V$ curves for a similar junction for resistances between 24
and 96 k$\Omega$, showing a fluctuation dominated Josephson
effect. Inset: representative $I-V$ data (symbols) and fit
(line).}
\end{figure}

Superconducting tips are prepared according to
Ref.~\onlinecite{Naaman01a} by deposition of Pb(5000
\nolinebreak\AA)/Ag(30 \nolinebreak\AA) proximity bilayers on
conventional PtIr tips. Fig.~\ref{fig1}(a) shows a typical dI/dV
curve for tunneling at 2.1 K between a SC tip and a Pb(5000
\nolinebreak \AA)/Ag(30 \nolinebreak\AA) sample deposited on a
graphite substrate at the same time as the tips were made. Sharp
peaks are seen at voltages $eV=\pm2\Delta_{Pb}$, and the Pb phonon
structure is clearly observed, confirming the high quality of the
junction and that single-step tunneling is the dominant
conduction mechanism. At these thicknesses (30 \AA) the Ag
contributes very little to the junction characteristics. The
method described above for SC tip fabrication is robust and
reproducible, yielding over 80\% working tips, with virtually
indistinguishable SC properties.

The measurement of the Josephson effect in STM junctions is
challenging due to their small size and the associated high
junction resistance. The energy scale for the phase coupling of a
Josephson junction is the Josephson binding energy, $E_J=\Delta
R_Q/2R_N$, where $R_Q=h/4e^2\sim6.45$ k$\Omega$, $\Delta$ is the
SC gap, and $R_N$ is the normal state resistance of the junction.
In junctions for which $k_BT\geq E_J$, phase locking is destroyed
by thermal fluctuations, and no dc supercurrent can be observed.
Indeed, for Pb junctions with resistances $\sim$100 M$\Omega$,
the Josephson binding energy is $E_J/k_B\sim0.5$ mK, and we see no
trace of the dc Josephson effect at our base temperature of 2.1 K
(Fig.~\ref{fig1}(a)).

Several authors have shown\cite{Ivanchenko69,Ambegaokar69} that
when $E_J$ is smaller than, but comparable to $k_BT$, the phase
dynamics in a small Josephson junction is dominated by thermal
fluctuations. In this regime pair tunneling would be observed,
but the pair current will be dissipative, with the voltage drop
proportional to the rate of change of the relative phase across
the junction. Ivanchenko and Zil'berman\cite{Ivanchenko69} found
that the current-voltage ($I-V$) relation for small $E_J/k_BT$
due to this incoherent pair tunneling has the form:
\begin{equation}
I(V)=\frac{I_c^2Z_{env}}{2}\frac{V}{V^2+V_p^2}; \label{IV}
\end{equation}
where $I_c=2eE_J/\hbar$, $Z_{env}$ is the effective impedance of
the junction's environment, and $V_p=(2e/\hbar)Z_{env}k_BT_n$
with $T_n$ defined as the effective noise temperature.

Fig.~\ref{fig1}(b) shows representative $I-V$ curves taken on a
Pb/Ag film at 2.1 K for various junction resistances\cite{note1}
below 120 k$\Omega$. A current peak in the $I-V$ characteristics
emerges from the quasiparticle background as the junction
resistance is lowered. The $I-V$ curves can be fitted with good
agreement to Eq.~\ref{IV} for $|V|<$1.0 mV, with $A\equiv
I_c^2Z_{env}/2$ and $V_p$ as the only fitting parameters. The
dependence of the quantity $\sqrt{(4e/\hbar)A/V_p}$ on
$G_N=1/R_N$ should be linear with a slope equal to
$I_cR_N/\sqrt{k_BT_n}$. Fig.~\ref{fig2} shows such a plot for
several Pb/Ag samples and a NbSe$_2$ sample. Knowing the noise
temperature $T_n$, one can infer the Josephson product $I_cR_N$.

Fluctuation dominated dc and ac Josephson effects were observed
and carefully analyzed in Ref.~\onlinecite{Naaman01b}, confirming
that the features we observe in the tunneling characteristics of
these STM junctions are indeed signatures of pair tunneling. We
have shown\cite{Naaman01b} that the noise parameters $T_n$ and
$Z_{env}$ depend only on the experimental setup and not on the
sample studied or the tip used. Once these parameters are
determined by measuring a sample with a known $I_cR_N$ ({\it
e.g.~}Pb), this quantity can be inferred for any sample measured
under the same experimental conditions\cite{note4}.

\section{Results and Discussion -- N\lowercase{b}S\lowercase{e}$_2$}

We apply the method described above to the measurement of the
Josephson product for tunneling between a SC STM tip and a
2H--NbSe$_2$ crystal. 2H--NbSe$_2$ is a layered material that
undergoes a charge density wave (CDW) transition at 33 K and a
superconducting transition at 7.2 K. Niobium diselenide has been
studied extensively in the
past\cite{Hess90,Truscott99,Yokoya01,Dordevic01,HudsonPhD}, but
to our knowledge, no reliable Josephson tunneling data exist for
this system. While NbSe$_2$ appears to be a conventional BCS
superconductor, it shows anisotropy of the SC order
parameter\cite{HudsonPhD}, transport and optical
properties\cite{Dordevic01}, as well as coexisting CDW and SC
states. These make NbSe$_2$ an interesting system, especially
since the high-T$_c$ cuprates are believed to share some of these
properties\cite{Hoffman02}.

\begin{figure}[b]
\epsfxsize=\columnwidth \epsfysize=4.0in
\epsfbox{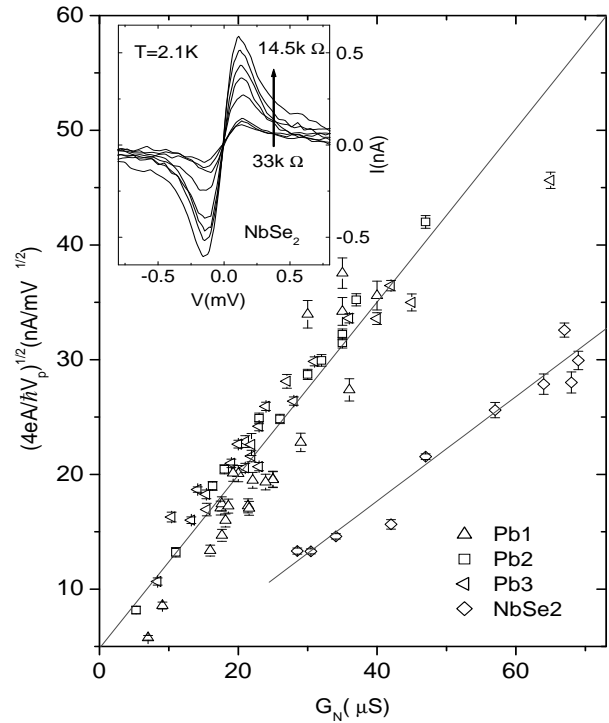} \caption{\label{fig2}Plot of
$\sqrt{(4e/\hbar)A/V_p}$ vs. $G_N$ for three Pb/Ag samples (Pb1,
Pb2, Pb3) and a NbSe$_2$ sample (diamond symbols). Solid lines
are linear fits to the data. The error in $G_N$ is estimated at
5\%. Inset: pair current vs. voltage for the NbSe$_2$ sample, and
junction resistances between 33 and 14.5 k$\Omega$.}
\end{figure}

$I-V$ curves were measured at 2.1 K on a 2H--NbSe$_2$ single
crystal\cite{Oglesby94} cleaved at room temperature in vacuum
($5\times10^{-7}$ torr). For junction resistances larger than 40
k$\Omega$, the scaled tunneling current $I(V)\times R_N$ is
independent of $R_N$. We use these curves to represent the
background quasiparticle current, with no Josephson contribution,
which we then subtract from the $I-V$ curves measured at lower
$R_N$ to find the contribution of pair tunneling to the total
current. This is done because junctions with NbSe$_2$ as one of
the electrodes have considerable quasiparticle weight below the
gap even at this temperature of 2.1 K.

Fig.~\ref{fig2} shows a plot constructed from the fits to
Eq.~\ref{IV} for several Pb/Ag films and the NbSe$_2$ data shown
in the inset. From linear fits to the data we obtain the slopes
$0.75\pm 0.03$ mV$^{1/2}$ and $0.46\pm 0.03$ mV$^{1/2}$ for Pb/Ag
and NbSe$_2$ respectively. The ratio of $I_cR_N$(Pb/I/NbSe$_2$)
and $I_cR_N$(Pb/I/Pb) is thus $0.61\pm 0.05$. Since
$I_cR_N$(Pb/I/Pb)=1.671 mV is known\cite{note2} from the
Ambegaokar-Baratoff (AB) formula\cite{Ambegaokar63}, we get
$I_cR_N$(Pb/I/NbSe$_2$)=$1.02\pm 0.08$ mV.

The SC gap in NbSe$_2$ can be described quite well by an
anisotropic s-wave gap function\cite{HudsonPhD}, with $\Delta(0)$
ranging from 0.7 to 1.4 meV over the Fermi surface (FS).
Estimates using the AB formula for Pb/I/NbSe$_2$ junctions yield
$I_c^{AB}R_N\sim$1.25 \nolinebreak mV, significantly higher than
the one we observe, even when using the smallest gap parameter in
NbSe$_2$. This suggests that the gap anisotropy alone cannot
explain our result and we must look for additional possible
explanations.

A further reduction in $I_cR_N$ may result from the existence of a
CDW and its effect on the SC state. Gabovich {\it et
al.~}\cite{Gabovich90} calculated the Josephson product for CDW
and SDW superconductors and found that it decreases from the AB
limit by an amount that depends on the ratio between the CDW and
the SC energy gaps, and the degree to which the FS is affected by
the formation of a CDW gap. In light of our results it appears
that explanations along these lines should be sought, especially
since the effects of a CDW may be important in the interpretation
of Josephson data in high-T$_c$ cuprates.

In summary, we measured the Josephson effect in junctions formed
between SC STM tips and Pb and NbSe$_2$ samples. In NbSe$_2$ we
find the Josephson product $I_cR_N=1.02\pm 0.08$ mV. This result
is smaller than expected from the AB formula, but may be
described within the theory of Gabovich {\it et al.~}for
Josephson tunneling in CDW superconductors. These results suggest
that the interplay between the SC state and the CDW may have an
important role in Josephson tunneling.

\begin{acknowledgments}
We would like to thank L. Bokacheva, W. Teizer, and the UCSD
Physics Electronics Shop. This work was supported by DOE Grant
number DE-FG03-00ER45853.
\end{acknowledgments}

\end{document}